\documentclass[12pt, a4paper]{article}

\usepackage[utf8]{inputenc}
\usepackage{geometry}
\usepackage{amsmath, amssymb}
\usepackage{authblk}
\usepackage{longtable}
\usepackage{caption}
\usepackage{etoolbox}
\usepackage{enumitem}
\usepackage{float}

\makeatletter
\AtBeginEnvironment{itemize}{\small\@minipagetrue}
\makeatother

\usepackage{graphicx}
\graphicspath{{images/}} 

\usepackage[authoryear,round]{natbib}
\usepackage{xcolor}
\usepackage[authoryear]{natbib}
\usepackage{hyperref}
\usepackage[capitalize,nameinlink]{cleveref} 
\crefname{figure}{Figure}{Figures}

\hypersetup{
    colorlinks=true,
    citecolor=blue,      
    linkcolor=blue,      
    urlcolor=blue,
}

\makeatletter
\renewcommand\citet[1]{%
  {\hypersetup{citecolor=black}\citeauthor{#1}} (\hyperlink{cite.#1}{\color{blue}\citeyear{#1}})%
}
\makeatother

\crefname{equation}{Eq.}{Eqs.}

\title{\textbf{From Patterns to Policy: A Scoping Review Based on Bibliometric Analysis (ScoRBA) of Intelligent and Secure Smart Hospital Ecosystems}}

\author[1*]{Adi Wijaya}
\author[2]{Budi Hermawan}
\author[3]{Wiga Maulana Baihaqi}
\author[4]{Catur Supriyanto}

\affil[1]{Department of Health Information Management, Universitas Indonesia Maju, Jakarta}
\affil[2]{Metrics Research Institute and Statistics Consulting, Indonesia}
\affil[3]{Department of Information Technology, Universitas Amikom Purwokerto, Purwokerto}
\affil[3]{Department of Electrical Engineering and Information Technology, Universitas Gadjah Mada, Yogyakarta}
\affil[4]{Faculty of Computer Science, Universitas Dian Nuswantoro, Semarang \vspace{0.2cm}}

\affil[ ]{\vspace{0.3cm} *Corresponding author: adiwjj@uima.ac.id}

\date{} 

\begin{document}

\maketitle
\vspace{-1.3cm}

\begin{abstract}
\noindent
This study examines the evolution of Intelligent and Secure Smart Hospital Ecosystems by applying a Scoping Review based on Bibliometric Analysis (ScoRBA) to systematically map research patterns, identify gaps, and derive policy implications. Using the Scopus database, a total of 891 journal articles published between 2006 and 2025 were analyzed through co-occurrence analysis, network visualization, overlay analysis, and the Enhanced Strategic Diagram (ESD). The analytical process is structured using the PAGER framework to link Patterns and Advances, Gaps and Research directions, and Evidence based policy implications.  The findings reveal three interrelated clusters that define the domain, namely AI driven intelligent healthcare systems, decentralized and privacy preserving digital health ecosystems, and scalable cloud edge infrastructures. These clusters indicate a convergence toward integrated, ecosystem level architectures where intelligence, trust, and infrastructure function as mutually reinforcing components. Despite significant advances in areas such as artificial intelligence, blockchain, and cloud computing, the study identifies key gaps related to interoperability, real world implementation, governance, and cross layer integration. Emerging themes such as explainable AI, federated learning, and privacy preserving mechanisms highlight underexplored areas that require further investigation.  Building on these insights, the study formulates policy relevant recommendations for both hospital management and government stakeholders, with particular attention to developing country contexts. The results emphasize the need for coordinated governance frameworks, scalable infrastructure strategies, and secure data ecosystems to ensure sustainable and effective digital health transformation. This study contributes by bridging bibliometric evidence with actionable policy directions, supporting the transition from technological patterns to informed policy making in smart hospital development.

\vspace{0.3cm} 
\end{abstract}

\section{Introduction}

Intelligent and secure smart hospital ecosystems are a revolutionary concept in healthcare as they combine different technologies such as IoT, AI, blockchain, and 5G/6G networks to improve patient care and hospital operations \citep{chow2025, kumar2025}. Such systems are designed to establish a fully integrated and data-driven environment in hospitals that will help in various aspects such as remote monitoring and diagnostics as well as administrative processes. One of the major reasons for the development of these systems is to offer personalized, proactive, and efficient healthcare services to cater the increasing number of elderly people globally and at the same time make healthcare infrastructures resilient \citep{bok2025}. But the downside of this fusion of technologies is the emergence of intricate issues regarding security of data, privacy, and interoperability, so that solid security measures and complete architectural models are of utmost importance \citep{hamouda2025, yadegari2025}. It is of paramount importance to maintain the integrity, confidentiality, and availability of sensitive patient data in such interconnected systems to build trust and unlock the full potential of smart hospitals.

Previous studies have revealed the potential of different components and aspects of smart hospital ecosystems that are not only intelligent but also secure. Investigations in this area have been carried out from the perspectives of developing secure access control mechanisms using smart contracts \citep{abid2024}, developing cyber engineering solutions that are resilient to attacks for IoT enabled smart healthcare systems \citep{pandey2024}, and adopting blockchain technology for securing and protecting privacy of data in IoT based systems \citep{alkhatib2025}. Moreover, the conceptualization and implementation of artificial intelligence and digital ecosystems for data-driven healthcare excellence in smart hospitals \citep{chow2025}, along with the consolidated proposal of IoT architectural models to enhance interoperability and efficiency \citep{yadegari2025}, are examples of efforts aimed at realizing advanced and secure healthcare environments through technical innovations. Only a small number of secondary studies, such as bibliometric and systematic reviews, have been carried out with the intention to explore this domain, yet their coverage is limited in that they tend to focus on particular domains or perspectives thereby leaving the broader thematic relationships relatively unexplored \citep{kaur2023, rasoulian2023, ozdemir2025}. Besides that, primary as well as review based studies are mostly focused on certain technologies or applications and give limited attention to integrated, system-level interpretations that would link technical, organizational, and governance dimensions together \citep{rabiei2025, loria2024}.

Despite the extensive body of research comprising primary investigations and secondary analyses, a large gap remains in the systematic detection of new and less explored topics in the intelligent and secure smart hospital ecosystems. Furthermore, many studies focus on individual aspects in isolation, which is common in the smart hospital literature, making it difficult to obtain a comprehensive understanding of thematic development and future directions \citep{malekzadeh2025}. Besides, there is limited translation of technical findings into policy and governance measures, which is a key part of the successful deployment and control of such systems. Hence, the absence of a thorough and systematically grouped synopsis prevents the stakeholders including policymakers, hospital management, and technology developers from making well-grounded decisions and setting up solid frameworks. Therefore, a Scoping Review based on Bibliometric Analysis (ScoRBA) can serve as an appropriate approach, as it combines the quantitative capability of bibliometric mapping to identify trends and patterns with qualitative synthesis to interpret thematic development and practical implications \citep{wijaya2023}. Accordingly, this study is guided by the following research questions (RQs):

\begin{enumerate}[label=\textbf{RQ\arabic*:}, leftmargin=*]
    \item What are the dominant patterns and key advances in the literature on Intelligent and Secure Smart Hospital Ecosystems?
    \item What underexplored areas exist in the literature on Intelligent and Secure Smart Hospital Ecosystems, and how can these areas shape a coherent future research agenda?
    \item What policy and practical implications can be derived from the existing evidence on Intelligent and Secure Smart Hospital Ecosystems to support smart hospital governance and digital health policy?
\end{enumerate}

The main aim of this research is to carry out a thorough mapping of the present situation of intelligent and secure smart hospital ecosystems. This aims to provide a structured understanding of existing developments, ongoing trends, and areas requiring further investigation. Using a ScoRBA approach, this study explores the interconnections between technological advancements, security considerations, and applications in smart hospitals through a comprehensive analysis. Besides, the study aims at translating the identified technical findings into policy and governance recommendations, thus providing stakeholders with useful guidance for building regulatory frameworks and best practices. In the end, this contributes to a more structured and informed approach to developing resilient, ethical, and secure smart healthcare environments, which would be in line with the future deployment of 6G and other advanced technologies in healthcare \citep{kumar2025}.

\section{Method}

This research adopts a Scoping Review based on Bibliometric Analysis (ScoRBA) method for a comprehensive mapping, analysis, and synthesis of the Intelligent and Secure Smart Hospital Ecosystems literature. The ScoRBA method, as described by \citet{wijaya2023}, is a combination of scoping review and bibliometric analysis that can be used as a unified methodological framework. The scoping review part uses the methods of \citet{arksey2005} as a structured guide for the identification, selection, and mapping of studies, whereas the bibliometric analysis part uses well established methods for the performance analysis and science mapping \citep{donthu2021}. Besides, the presentation and analysis of findings are done according to the PAGER framework \citep{bradburyjones2022}. As a result of bringing together these complementary guidelines, ScoRBA allows a comprehensive exploration of research patterns, gaps, and implications. This method is particularly suitable for complicated and multidisciplinary fields as it combines quantitative bibliometric mapping with qualitative interpretation so the research can go beyond mere description to structured knowledge development and policy relevant insights.

\subsection{Data Collection}

\begin{figure}[htbp]
    \centering
    \includegraphics[width=0.9\textwidth]{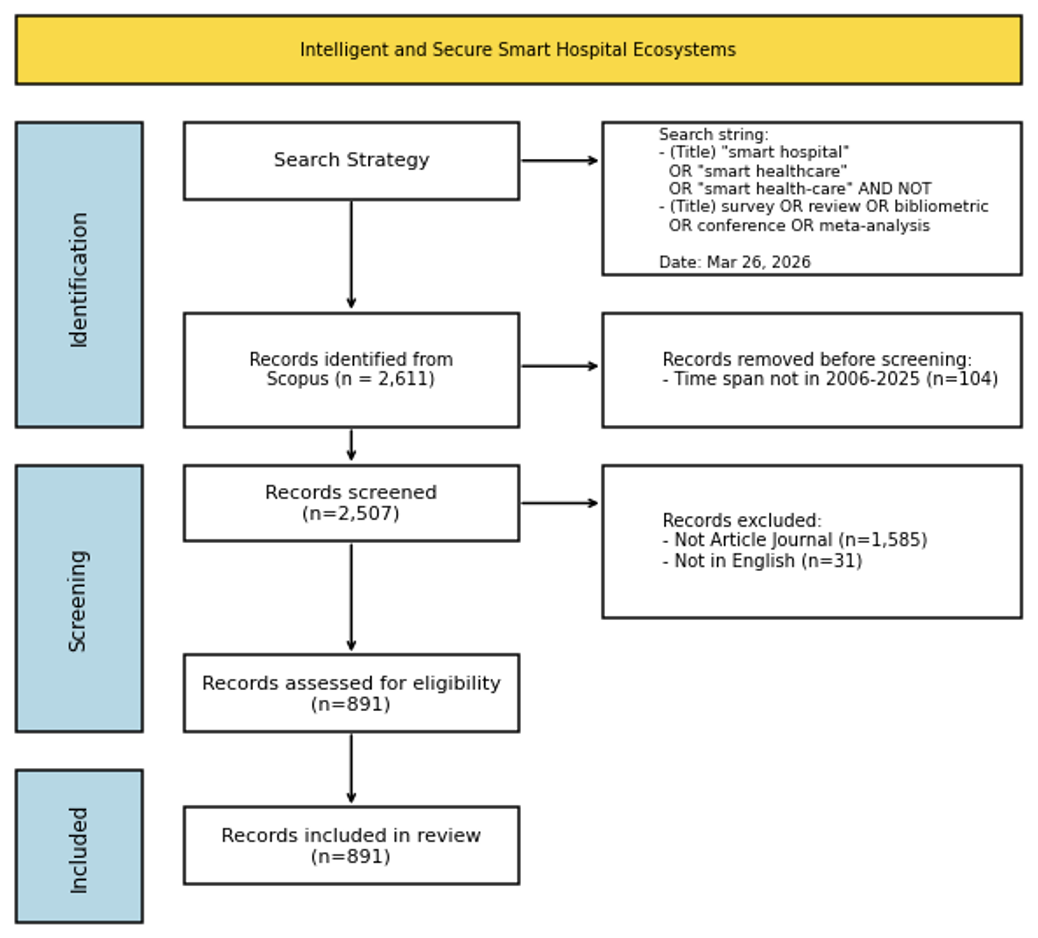} 
    \caption{Search strategy adapted from the PRISMA flow diagram \citep{page2021}}
    \label{fig:prisma_flow}
\end{figure}

The data collection process follows a transparent and replicable protocol guided by the PRISMA flow diagram, adapted from established reporting standards \citep{page2021}, as illustrated in \cref{fig:prisma_flow}. The search strategy was conducted exclusively in the Scopus database using combinations of keywords such as smart hospital, smart healthcare, and smart health-care, while excluding review-based studies (e.g., survey, review, bibliometric, and meta-analysis) to ensure the inclusion of primary research contributions. This initial search resulted in 2,611 records, which were subsequently filtered based on a predefined publication time span (2006–2025), leading to the removal of 104 records outside the specified range.

The screening phase did not involve title or abstract screening, but instead applied strict filtering criteria based on document type and language, retaining only journal articles and English-language publications. Altogether, there were 891 records that after this process were considered eligible and included in the final analysis. The exclusive use of the Scopus database is a well-grounded decision as this database is characterized by extensive coverage and rich citation data, as it contains more publications and citation records than other databases, thereby allowing for a more thorough representation of the scientific landscape \citep{zhu2020, pranckute2021}. Such comprehensive coverage is a prerequisite for accurately detecting research trends, thematic structures, and emerging topics in the field \citep{zakaria2022}.

\subsection{Data Analysis}

\begin{figure}[htbp]
    \centering
    \includegraphics[width=0.8\textwidth]{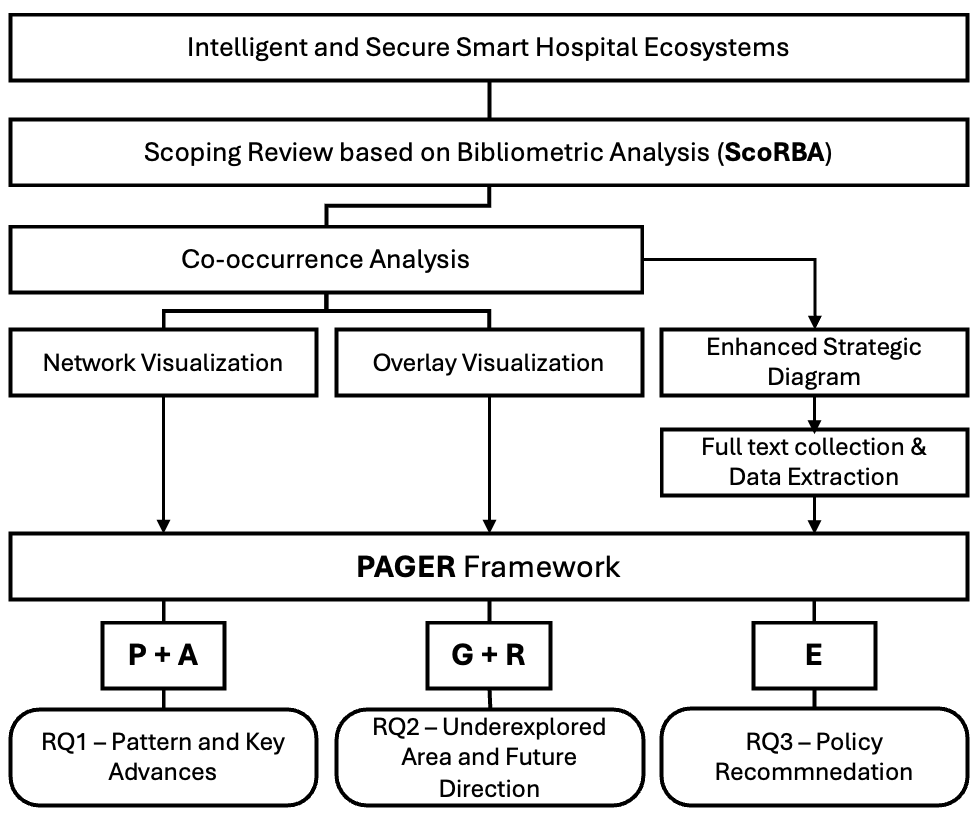} 
    \caption{Data analysis processes}
    \label{fig:data_analysis}
\end{figure}

The data analysis process is structured based on the ScoRBA framework, as depicted in \cref{fig:data_analysis}, which integrates co-occurrence analysis, network visualization, overlay visualization, and the Enhanced Strategic Diagram (ESD). Co-occurrence analysis is employed to identify relationships among keywords, enabling the formation of thematic clusters that represent the intellectual structure of the research field. These clusters are further examined using network visualization to uncover the structural interconnections among topics, while overlay visualization highlights temporal dynamics, particularly the emergence of recent and high-impact research areas.

To enhance the analysis, this paper incorporates the Enhanced Strategic Diagram (ESD) as an extension of conventional bibliometric mapping methods, offering a more comprehensive, multidimensional categorization of research themes \citep{shafin2022}. The ESD categorizes keywords into eight quadrants depending on their total link strength (x-axis) and occurrence (y-axis), together with the temporal differentiation between recent and older publications. In this way, it enables the identification of core, emerging, mature, and isolated themes, thus providing a broader perspective of dominant as well as less explored areas. The combination of these analytical methods guarantees that the depiction of the research terrain is not only structural but also temporal and strategic.

The outputs of the bibliometric analysis are subsequently synthesized using the PAGER framework, where Patterns and Advances (P+A) are used to address RQ1, Gaps and Research recommendation (G+R) inform RQ2, and Evidence for practice (E) supports the derivation of policy implications in RQ3. This integrated analytical flow enables a coherent transition from data-driven mapping to conceptual synthesis and practical recommendations, ensuring that the findings are both methodologically robust and contextually relevant, particularly for advancing smart hospital ecosystems in diverse and developing settings.

\section{Results and Discussion}

\subsection{Descriptive Results}

The descriptive analysis offers a preliminary overview of the publication trends, major contributors, as well as thematic changes in the field of Intelligent and Secure Smart Hospital Ecosystems. This part mainly illustrates the timeline of the research area's evolution, points out the significant authors and sources who have been instrumental in shaping the scholarly discourse, and records the temporal emergence of leading topics. As a whole, these insights establish the empirical foundation for understanding the patterns, progress, and research paths that are elaborated on in the following sections.

\begin{figure}[htbp]
    \centering
    \includegraphics[width=0.85\textwidth]{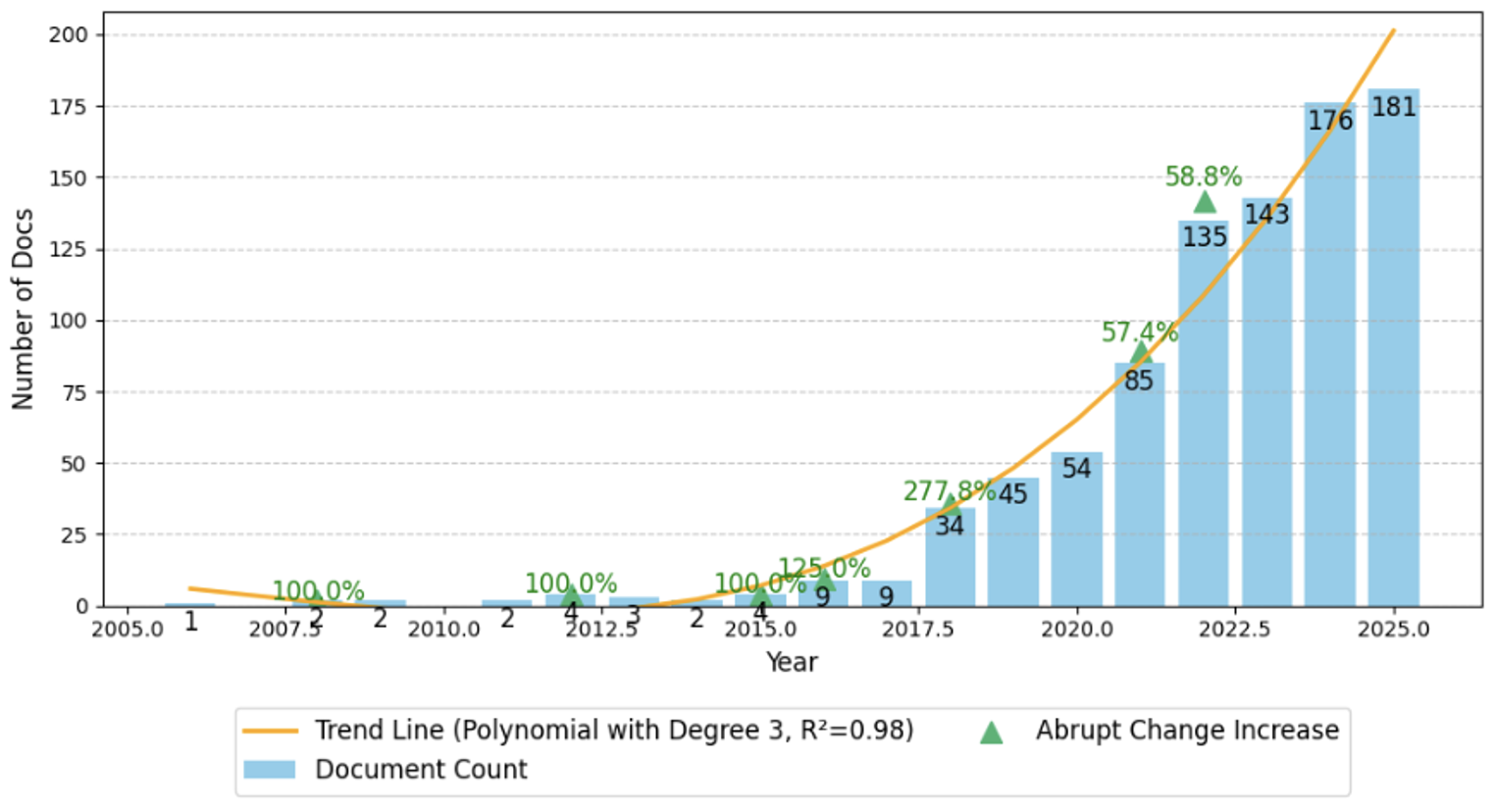} 
    \caption{Publication over years}
    \label{fig:pub_years}
\end{figure}

As shown in \cref{fig:pub_years}, the number of publications shows almost a perfect continual increase, especially after 2018. In fact, the earlier years display very limited activity as there were less than five publications annually during 2006--2014 implying that the research was just beginning. Then, there was a significant increase from 2018 when publications rose substantially from 34 in 2018 to 85 in 2021, and further to 181 in 2025. The polynomial trend and a high $R^2$ value of 0.98 point to a very reliable and continued growth pattern which means that the subject has become a well established and very fast developing research field. Besides, the sharp times of growth around 2018 and 2021--2022 show that many researchers have been attracted and want to know more due to the quite a number of new developments in AI, IoT, and digital health transformation.

\begin{figure}[htbp]
    \centering
    \includegraphics[width=1\textwidth]{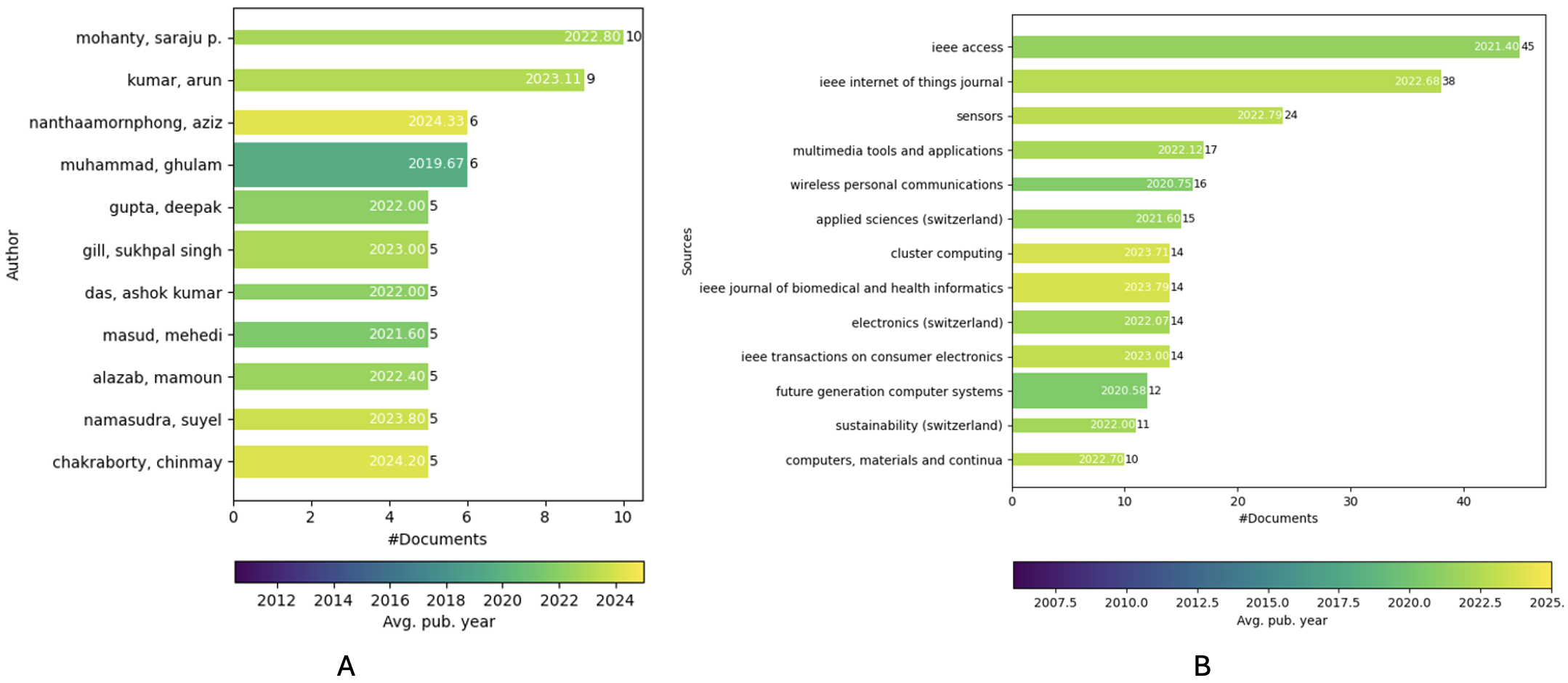} 
    \caption{Prolific and influential authors (A) and sources (B)}
    \label{fig:authors_sources}
\end{figure}

\hyperref[fig:authors_sources]{Figure \ref{fig:authors_sources}A} presents the top authors who publish the most and have the highest impact in the field, revealing a mix of well known and rising authors. Kumar and Gill, for example, have the highest average normalized citations, which means their research in the last few years has had a strong impact. Notably, Kumar and Gill have authored papers on next generation smart hospital infrastructures, AIoT integration, and latency optimization in healthcare systems enabled by 5G and 6G \citep{kumar2020, tuli2020}. On the other hand, Muhammad, although has an earlier average publication year, remains highly cited mainly because of his participation in the most cited federated learning and AI based healthcare frameworks papers, which suggests the perennial importance of foundational work \citep{rahman2023}. This is what happens with other authors such as Mohanty and Das, whose prior work on edge computing, security, and IoT based healthcare are still actively cited \citep{sayeed2019, chaudhary2018}. In contrast, some of the latest papers from authors like Nanthaamornphong and Chakraborty have been highly cited for topics like 6G enabled healthcare connectivity and AI based medical imaging \citep{kumar2025, mazumdar2025}. The variation in bar color and width further reflects the balance between recency and impact, while the keyword distribution dominated by themes such as smart healthcare, IoT, AI, security, and blockchain highlights a converging research focus on intelligent, secure, and connected healthcare ecosystems, underscoring a dynamic and rapidly evolving authorship landscape.

As shown in \hyperref[fig:authors_sources]{Figure \ref{fig:authors_sources}B}, most productive and influential sources indicate that journals like IEEE Access and IEEE Internet of Things Journal are leading in terms of number of publications, which is indicative of their main roles in publishing research in this domain. On the other hand, journals like IEEE Journal of Biomedical and Health Informatics and Future Generation Computer Systems have higher average normalized citations reflecting their superior influence in the growth of research. Besides, more recent sources such as Cluster Computing and Electronics (Switzerland) have relatively new average publication years, implying their increasing importance. This pattern reveals that while prominent journals still act as the major publication venues, new journals are progressively providing high impact research.

\begin{figure}[htbp]
    \centering
    \includegraphics[width=0.8\textwidth]{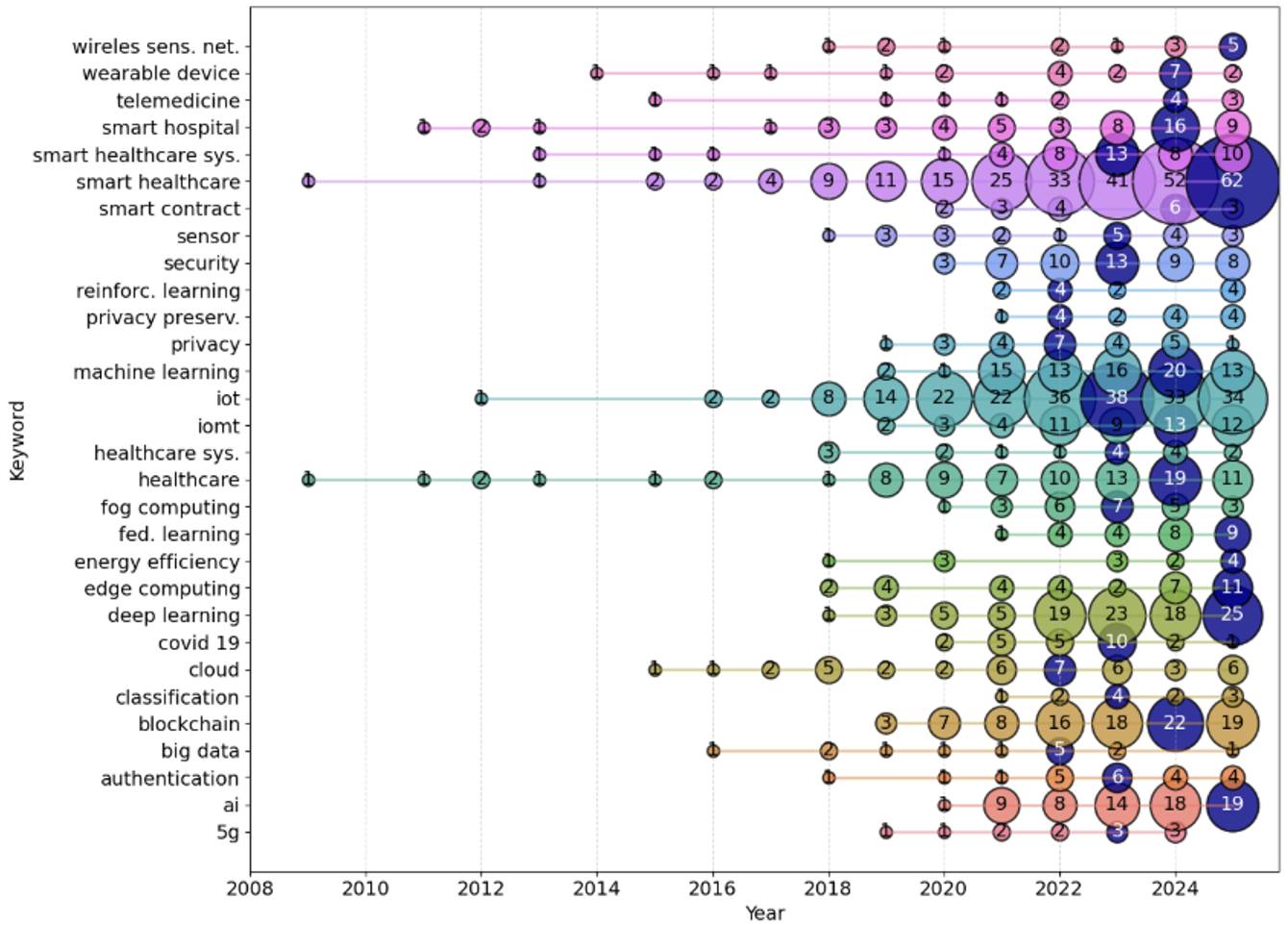} 
    \caption{Topics occurrence over years}
    \label{fig:topics_years}
\end{figure}

The temporal evolution of research topics is described in \cref{fig:topics_years}. It shows both the current main and future topics of the field. Most of the main topics, for example smart healthcare, IoT, and deep learning, have the highest number of occurrences in the past few years and are expected to continue to peak in 2024 and 2025, which confirms their pivotal position in the domain. Other technologies like blockchain, machine learning, and AI have shown strong and steady growth, which corresponds to their incorporation into the smart hospital ecosystem. New topics such as federated learning, edge computing, wireless sensor networks, and energy efficiency have more recent peaks and therefore tend to be the ones that are aligned with the ideas of privacy preservation, distributed, and sustainable system architectures, while also promoting the trust and protection mechanisms in conjunction with the technological progress as suggested by the presence of the words like security, IoMT, and privacy.

These descriptive results collectively indicate a research field that is not only rapidly expanding but also evolving toward greater integration across intelligence, security, and infrastructure dimensions. The convergence of high frequency topics with emerging trends suggests a transition from foundational technological exploration toward more complex, system level implementations, where scalability, interoperability, and governance become increasingly critical.

\subsection{RQ1 - Patterns and Key Advances}

\begin{figure}[htbp]
    \centering
    \includegraphics[width=0.7\textwidth]{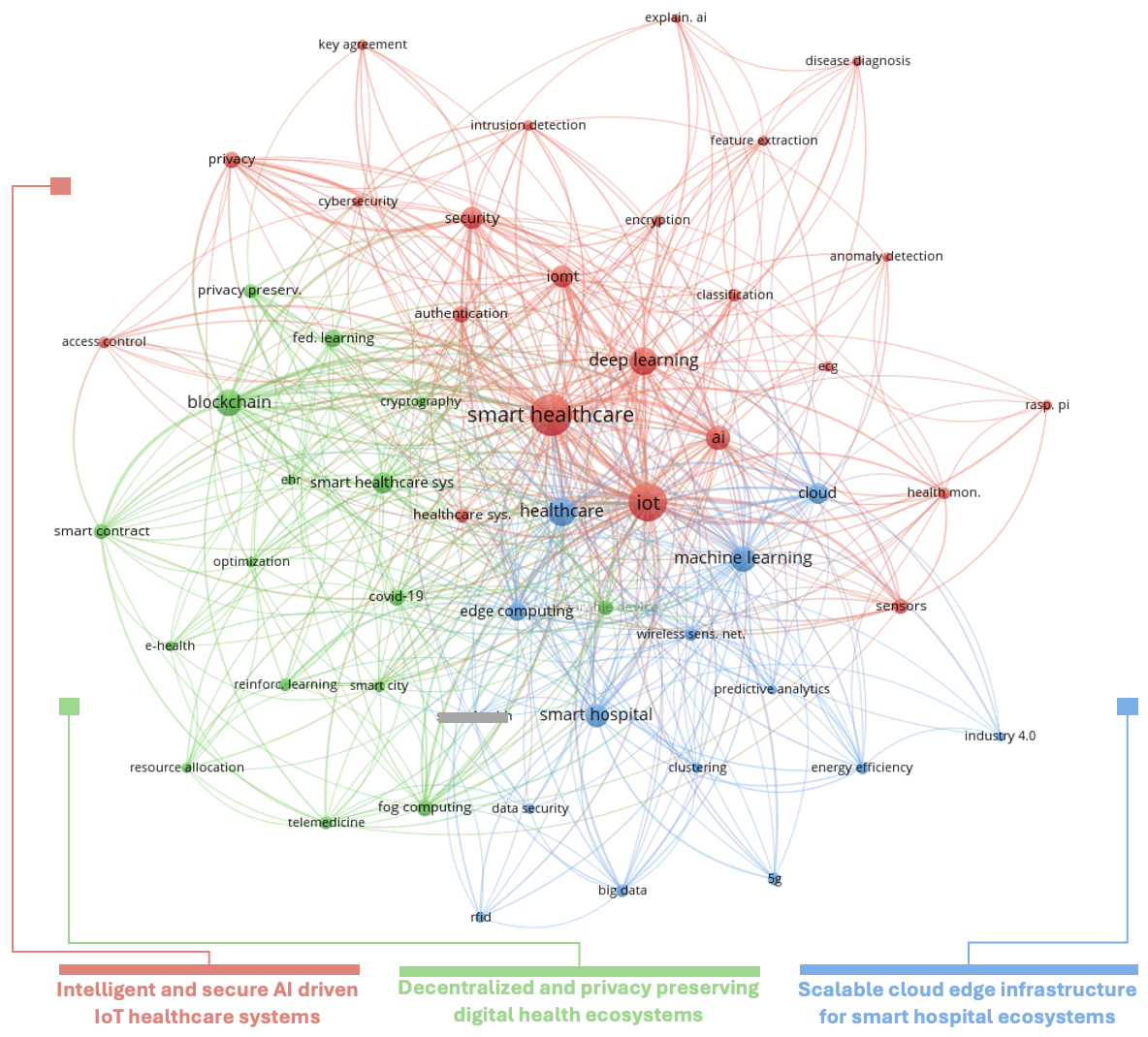} \\
    \vspace{0.2cm}
    \textbf{A} 
    
    \vspace{0.3cm} 
    
    \includegraphics[width=0.6\textwidth]{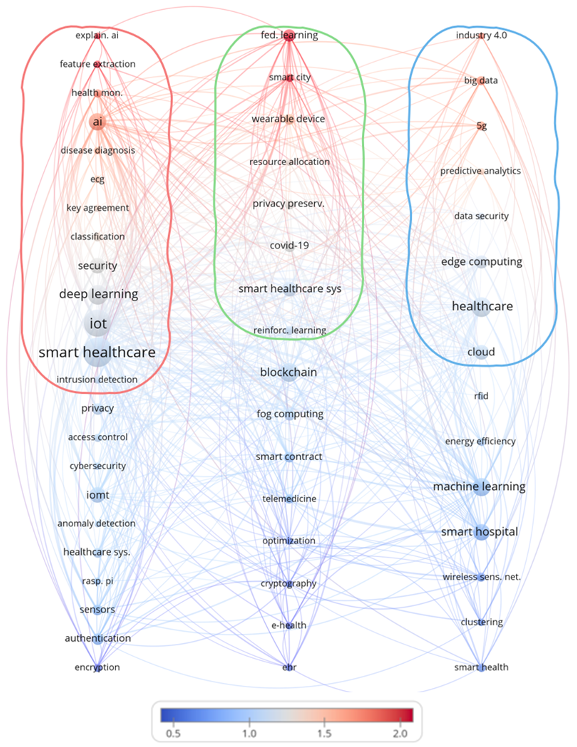} \\
    \vspace{0.2cm}
    \textbf{B}
    
    \caption{Network visualization with General Theme (A) and selected topics (B) for each cluster}
    \label{fig:network_viz}
\end{figure}

The bibliometric structure indicates three main overlapping thematic clusters that together explain the development of Intelligent and Secure Smart Hospital Ecosystems, which cover intelligence, trust, and infrastructure layers, as seen in \hyperref[fig:network_viz]{Figure \ref{fig:network_viz}A}. The study with a tightened criteria of normalized citation ($\geq 1.1$) manages to trace not only the mainstream but also the highly influential research streams that bring a good combination of representativeness and academic impact, as illustrated in \hyperref[fig:network_viz]{Figure \ref{fig:network_viz}B} and Table \cref{tab:pattern_advances}. Although the clusters are still linked to each other by main connecting nodes like AI, IoT, and healthcare systems, they mirror a unification towards fully integrated, intelligent, and secure digital healthcare environments. Such an arrangement reveals the development from isolated technological usage to the level of system orchestration where intelligence, trust, and infrastructure are evolving together, including smart healthcare systems enabled by AI and IoT \citep{khanh2025, almufareh2025, hermawan2022}, cloud computing and big data infrastructures \citep{gao2024, tuli2020}, as well as federated learning and privacy preserving mechanisms for secure healthcare data management \citep{rahman2023, ullah2024, akter2022, akter2024, lim2021}.

The first cluster relates to the concept of Intelligent and Secure AI-Driven IoT Healthcare Systems as the operational intelligence layer of smart hospitals. As clinical intelligence in the healthcare area is becoming more data-driven, a strong presence of AI, deep learning, and disease diagnosis illustrates the capability of technology to detect and forecast health-related conditions \citep{nancy2022, golec2023}. In addition to that, the use of ECG and health monitoring also indicates that continuously collected real-time patient data contributes significantly to improved clinical outcomes \citep{mubarakali2025, nancy2022}. Furthermore, the development of explainable AI and feature extraction reflects increased attention to interpretable and trustworthy AI systems, which is important for ensuring transparency and accountability in clinical decision-making \citep{mubarakali2025}. On the other hand, the integration of security, intrusion detection, and key agreement indicates that trust mechanisms are increasingly incorporated into intelligent systems, allowing secure operation in sensitive healthcare environments \citep{keshta2022}. Apart from this, the inclusion of IoT and smart healthcare highlights the progression toward connected and real-time healthcare ecosystems where intelligent analytics and physical devices operate together \citep{minopoulos2022, khanh2025, golec2023}. These developments indicate a shift toward integrated, real-time, and secure AI-driven healthcare systems.

\renewcommand{\arraystretch}{1.5} 

\begin{longtable}{|p{0.15\textwidth}|p{0.3\textwidth}|p{0.45\textwidth}|}
    \caption{Pattern and key advances for each cluster}
    \label{tab:pattern_advances} \\
    
    \hline
    \textbf{Cluster} & \textbf{Pattern (General Theme)} & \textbf{Key Advances} \\
    \hline
    \endfirsthead
    
    \multicolumn{3}{c}{{\bfseries \tablename\ \thetable{} -- continued from previous page}} \\
    \hline
    \textbf{Cluster} & \textbf{Pattern (General Theme)} & \textbf{Key Advances} \\
    \hline
    \endhead
    
    \hline \multicolumn{3}{|r|}{{Continued on next page...}} \\ \hline
    \endfoot
    
    \hline
    \endlastfoot

    Cluster 1 (Red) & Intelligent and Secure AI Driven IoT Healthcare Systems & 
    \begin{itemize}[nosep, leftmargin=*, before=\vspace{-0.6\baselineskip}, after=\vspace{0.5\baselineskip}]
        \item Development of AI-driven clinical intelligence for diagnosis and monitoring;
        \item Emergence of interpretable and trustworthy AI systems;
        \item Integration of security-aware intelligent systems;
        \item Advancement of real-time and connected healthcare systems
    \end{itemize} \\
    \hline
    
    Cluster 2 (Green) & Decentralized and Privacy Preserving Digital Health Ecosystems & 
    \begin{itemize}[nosep, leftmargin=*, before=\vspace{-0.6\baselineskip}, after=\vspace{0.5\baselineskip}]
        \item Emergence of privacy-preserving collaborative intelligence;
        \item Expansion of distributed and real-world healthcare applications;
        \item Development of adaptive and resource-aware healthcare systems;
        \item Strengthening of system-level decentralized healthcare architectures
    \end{itemize} \\
    \hline
    
    Cluster 3 (Blue) & Scalable Cloud Edge Infrastructure for Smart Hospital Ecosystems & 
    \begin{itemize}[nosep, leftmargin=*, before=\vspace{-0.6\baselineskip}, after=\vspace{0.5\baselineskip}]
        \item Advancement of data-driven healthcare intelligence at scale;
        \item Development of cloud-edge computing architectures;
        \item Emergence of high-speed and connected infrastructures;
        \item Integration of secure and scalable digital healthcare platforms;
        \item Alignment with Industry 4.0 paradigms for system automation
    \end{itemize} \\
\end{longtable}

The second cluster is about Decentralized and Privacy Preserving Digital Health Ecosystems, placing healthcare on the path toward distributed and trust-enabled architectures. The significant influence of federated learning and privacy measures indicates a shift toward collaborative intelligence models that safeguard sensitive data while enabling cross-institutional analytics \citep{akter2024, jiang2025, guo2024}. The recognition of wearable devices and COVID-19 suggests an increasing reliance on real-world and patient-oriented healthcare applications, especially in remote monitoring and pandemic response scenarios \citep{patel2024, ozturk2025}. On top of that, the reference to resource allocation and reinforcement learning shows that there is an ongoing trend of developing adaptive and optimized healthcare systems, which can effectively manage resources in distributed environments dynamically \citep{dong2024, sankaradass2025}. Smart city in turn helps healthcare become part of a broader digital ecosystem, highlighting cross-sector integration \citep{goel2024, ali2025}. These changes imply that the trust layer is being transformed into privacy-aware, distributed, and adaptive healthcare intelligence systems.

The third cluster relates to scalable cloud edge infrastructure for smart hospital ecosystems, which is essentially the fundamental framework that supports intelligent and trustworthy healthcare services. The prevalence of big data and predictive analytics underscores the role of extensive data processing and the ability to forecast as key factors in enhancing both clinical and operational decision making \citep{hoang2022, nancy2022}. The combination of cloud and edge computing points to the creation of versatile and decentralized computing structures, which are not only capable of handling real time but also high volume healthcare applications \citep{goyal2021, rajavel2022}. The introduction of 5G makes the whole infrastructure even stronger by providing very fast speed and very low latency connection, which is critical for real time healthcare services and the integration of IoT \citep{gupta2023, pradhan2023, tang2022}. Moreover, data security and healthcare systems' inclusion here signifies the growing importance of secure and interoperable digital platforms, whereas industry 4.0 stands for the merging with automation and system optimization based on very advanced technology \citep{aldosary2024, kumar2020}. Hence, these innovations present infrastructure as a scalable, secure, and high performance digital backbone.

The refined patterns and advances illustrate a progression toward integrated smart hospital ecosystems, where intelligence, trust, and infrastructure function as mutually reinforcing components. The convergence of AI-driven analytics, privacy-preserving distributed intelligence, and scalable cloud-edge infrastructures reflects a shift toward holistic system orchestration, where technological capabilities are increasingly aligned with requirements for security, interoperability, and real-time responsiveness \citep{mohammed2023, yadegari2025}.

\subsection{RQ2 - Underexplored Area and Future Research Directions}

\begin{figure}[!t]
    \centering
    \includegraphics[width=0.8\textwidth]{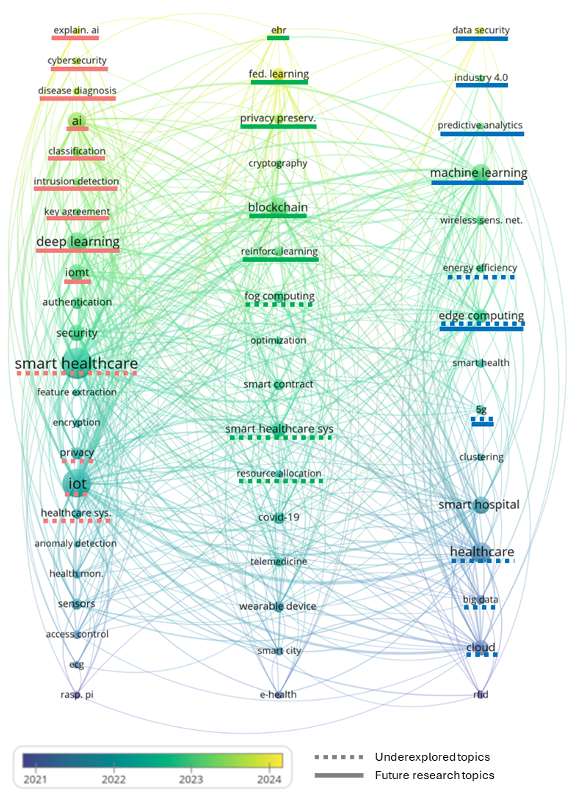} 
    \caption{Selected underexplored and future direction topics for each cluster}
    \label{fig:underexplored_topics}
\end{figure}

The literature review conducted with the help of time filtering shows that several issues that have yet to be investigated are expressed as structural deficiencies in the evolution of Intelligent and Secure Smart Hospital Ecosystems. Restricting to the 2021--2023 period, the data exposes that even though some fields are developing, they remain largely uncharted, especially in terms of seamless integration of intelligence, security, and infrastructure layers. These gaps are not only the result of partial implementations but also signify the fundamental constraints of how system level orchestration, interoperability, and real life applicability are addressed by the current research \citep{akram2024, talaat2022}. The distribution of these less researched topics throughout the different thematic clusters is depicted in \cref{fig:underexplored_topics}, thus offering a graphical representation of research gaps, whereas Table \cref{tab:underexplored_future} serves as a document where these gaps are systematically tied to the future research directions for each cluster. Meanwhile, papers published starting in 2023 outline a series of future investigations that revolve around a move toward better adaptive, easily understandable, and decentralized models \citep{chen2024, sarosh2021, selvaraj2024}. This temporal distinction, as reflected in \cref{fig:underexplored_topics} and elaborated in Table \cref{tab:underexplored_future}, provides a rigorous basis for separating what remains underexplored from what is emerging as the research frontier, enabling a coherent formulation of a future research agenda grounded in both unresolved challenges and advancing innovations.

The first cluster deals with intelligent and secure AI powered IoT healthcare systems. Here, the literature reveals that current advances are mostly system fragmented even though the supporting technologies are becoming mature. The integration of IoT, smart healthcare, healthcare system, and privacy lead to transitional gaps, as these elements are developed independently more often than as part of the unified clinical systems \citep{akter2022, liu2022}. Also, though AI driven techniques have been extensively studied, most of their application is still model centric and does not consider real world clinical workflows and operational constraints \citep{rattal2025}. Besides, security measures are still not completely integrated in the AI-IoMT lifecycle, which leads to disjointed networks in protection and trust management \citep{kumar2020, morovisconti2020}. On the other hand, the front line of research is changing the focus of clinically deployable and trustworthy AI systems. Explainable AI, disease diagnosis AI, classification, and deep learning are all being focused on making AI interpretable, accountable, and usable in the real world \citep{chen2024, khadidos2023, kumar2025, mansour2021}. In parallel, the advent of cybersecurity, intrusion detection, and key agreement points to a new security by design approach, where the safeguarding elements are built into the smart healthcare systems themselves \citep{amintoosi2022, ravi2024}. The addition of IoMT (Internet of Medical Things) is a further step toward the creation of comprehensive intelligent and secure healthcare ecosystems that combine analytics, connectivity, and trust within one operational framework \citep{kumar2025}.

For the second cluster which depicts decentralized and privacy preserving digital health ecosystems, the deficiencies especially relate to scalability, interoperability and governance aspects. Transitional traces indicate that technologies like smart healthcare systems, fog computing, and resource allocation emphasize system level problems, mainly in organizing distributed healthcare services and handling data flows across institutions \citep{moqurrab2022, talaat2022}. On the other hand, such methods are still largely disconnected, while governance factors such as data ownership, consent management, and institutional coordination have not reached a mature stage. The frontier research direction however reveals a strong inclination toward fully decentralized and privacy preserving intelligence ecosystems, which are facilitated by the integration of federated learning \citep{chen2024}, privacy preservation \citep{liu2022}, and blockchain \citep{xie2021}. These technologies together make it possible to conduct collaborative analytics without the need for direct data sharing, thereby solving the problems of both scalability and privacy \citep{agarwal2024, bai2022}. Transforming next generation Electronic Health Records (EHR) is a very important aspect. In this transformation, EHR will change from being a simple data storage to becoming a live, interoperable, and decentralized data platform \citep{faneela2023, szczepaniuk2023}. Also, with reinforcement learning, the future systems could be decentralized as well as adaptive \citep{dong2024, rattal2025}. These systems will optimize resource allocation and system behavior instantaneously over distributed healthcare environments.

The third cluster, which represents scalable cloud edge infrastructure for smart hospital ecosystems, points out the challenges in the healthcare domain related to the actual use of infrastructure capabilities. Recent papers show that technologies like edge computing, energy efficiency, and 5G are becoming very popular \citep{balasundaram2023, gharaei2024}, but their use in clinical workflows and healthcare systems is still very scarce. Even though there are basic pieces such as cloud and data platforms, the meeting point of infrastructure performance and healthcare specific needs, such as interoperability and real time responsiveness, is still very weakly tackled \citep{ji2024, mohammed2023}. On the other hand, frontier research is going towards the concept of intelligent and autonomous infrastructure ecosystems, which rely on data security, smart prediction, and industry 4.0 to deliver healthcare services that are real time, adaptive, and context aware \citep{kumar2020, nancy2022, sarosh2021}. Machine learning is becoming one of the main integration points, as a form of intelligence bridging different layers, whereby machine learning is at the top of the heap when it comes to predictive optimization, dynamic resource management, and automated orchestration of healthcare systems \citep{abdulkareem2021, nasr2021}. This transition is also being driven by the merging of edge computing, high speed connectivity, and AI which are all the elements of a smart hospital environment that is both responsive and scalable \citep{yadav2021}.

\renewcommand{\arraystretch}{1.5}

\begin{longtable}{|p{0.15\textwidth}|p{0.38\textwidth}|p{0.38\textwidth}|}
    \caption{Underexplored Area and Future Research Directions for each cluster}
    \label{tab:underexplored_future} \\
    
    \hline
    \textbf{Cluster} & \textbf{Underexplored area} & \textbf{Future Research Directions} \\
    \hline
    \endfirsthead
    
    \multicolumn{3}{c}{{\bfseries \tablename\ \thetable{} -- continued from previous page}} \\
    \hline
    \textbf{Cluster} & \textbf{Underexplored area} & \textbf{Future Research Directions} \\
    \hline
    \endhead
    
    \hline \multicolumn{3}{|r|}{{Continued on next page...}} \\ \hline
    \endfoot
    
    \hline
    \endlastfoot

    Cluster 1 (Red) & 
    \begin{itemize}[nosep, leftmargin=*, before=\vspace{-0.6\baselineskip}, after=\vspace{0pt}]
        \item Limited integration of intelligent and secure AI-driven systems within clinical workflows
        \item Predominance of model-centric AI development with insufficient system-level deployment and alignment with healthcare operations
        \item Fragmented and incomplete embedding of security mechanisms across AI--IoMT environments
    \end{itemize} & 
    \begin{itemize}[nosep, leftmargin=*, before=\vspace{-0.6\baselineskip}, after=\vspace{0pt}]
        \item Development of clinically deployable and explainable AI systems for real-world healthcare adoption
        \item Design of integrated and adaptive security architectures incorporating cybersecurity mechanisms within AI--IoMT ecosystems
        \item Exploration of end-to-end intelligent healthcare systems unifying analytics, security, and operational workflows
    \end{itemize} \\
    \hline

    Cluster 2 (Green) & 
    \begin{itemize}[nosep, leftmargin=*, before=\vspace{-0.6\baselineskip}, after=\vspace{0pt}]
        \item Limited interoperability and scalability in decentralized health systems, with insufficient integration of smart healthcare, fog computing, and resource allocation
        \item Persistent governance and coordination gaps in multi-institutional settings, including fragmented data sharing, consent management, and operational alignment
    \end{itemize} & 
    \begin{itemize}[nosep, leftmargin=*, before=\vspace{-0.6\baselineskip}, after=\vspace{0pt}]
        \item Advancement of privacy-preserving and decentralized intelligence ecosystems integrating federated learning, privacy preservation, and blockchain
        \item Redefinition of next-generation EHR systems as interoperable, decentralized, and privacy-aware platforms
        \item Exploration of adaptive decentralized systems using reinforcement learning for dynamic resource and system management
    \end{itemize} \\
    \hline

    Cluster 3 (Blue) & 
    \begin{itemize}[nosep, leftmargin=*, before=\vspace{-0.6\baselineskip}, after=\vspace{0pt}]
        \item Limited operationalization of scalable healthcare infrastructure, with insufficient integration of cloud, big data, and real-time intelligent systems
        \item Emerging capabilities (e.g., edge computing, energy efficiency, 5G) remain weakly aligned with clinical workflows, interoperability, and real-time responsiveness
    \end{itemize} & 
    \begin{itemize}[nosep, leftmargin=*, before=\vspace{-0.6\baselineskip}, after=\vspace{0pt}]
        \item Development of intelligent and autonomous infrastructure ecosystems enabling adaptive, real-time healthcare services
        \item Integration of machine learning for predictive optimization, dynamic resource management, and automated system orchestration
        \item Exploration of next-generation edge-enabled healthcare systems combining edge computing, 5G, and AI for scalable smart hospital environments
    \end{itemize} \\
\end{longtable}

These discoveries taken together imply future investigations should no longer focus on separate technological advances but on creating integrated, adaptive, and intelligent hospital ecosystems \citep{kavarthapu2024, priya2023}. Moreover, the existence of loopholes across all clusters clearly reveals that the global supply of enabling technologies is not the main issue anymore; rather, the major problem is their integration at the system level and alignment with operations. On the other hand, the directions of the frontier show a very evident path to explainable intelligence, decentralized trust, and autonomous infrastructure \citep{jiang2025, lamba2024, levina2024}. Working across these three sides will need a research program which is able to interpret, communicate, and expand at the same time so that technological innovation can be turned into health care service in the real world \citep{chen2024, akter2022, aldosary2024, nancy2022, rajavel2022, kumar2020}.

\subsection{RQ3 - Policy and Practical Implications}

\begin{figure}[H]
    \centering
    \includegraphics[width=0.75\textwidth]{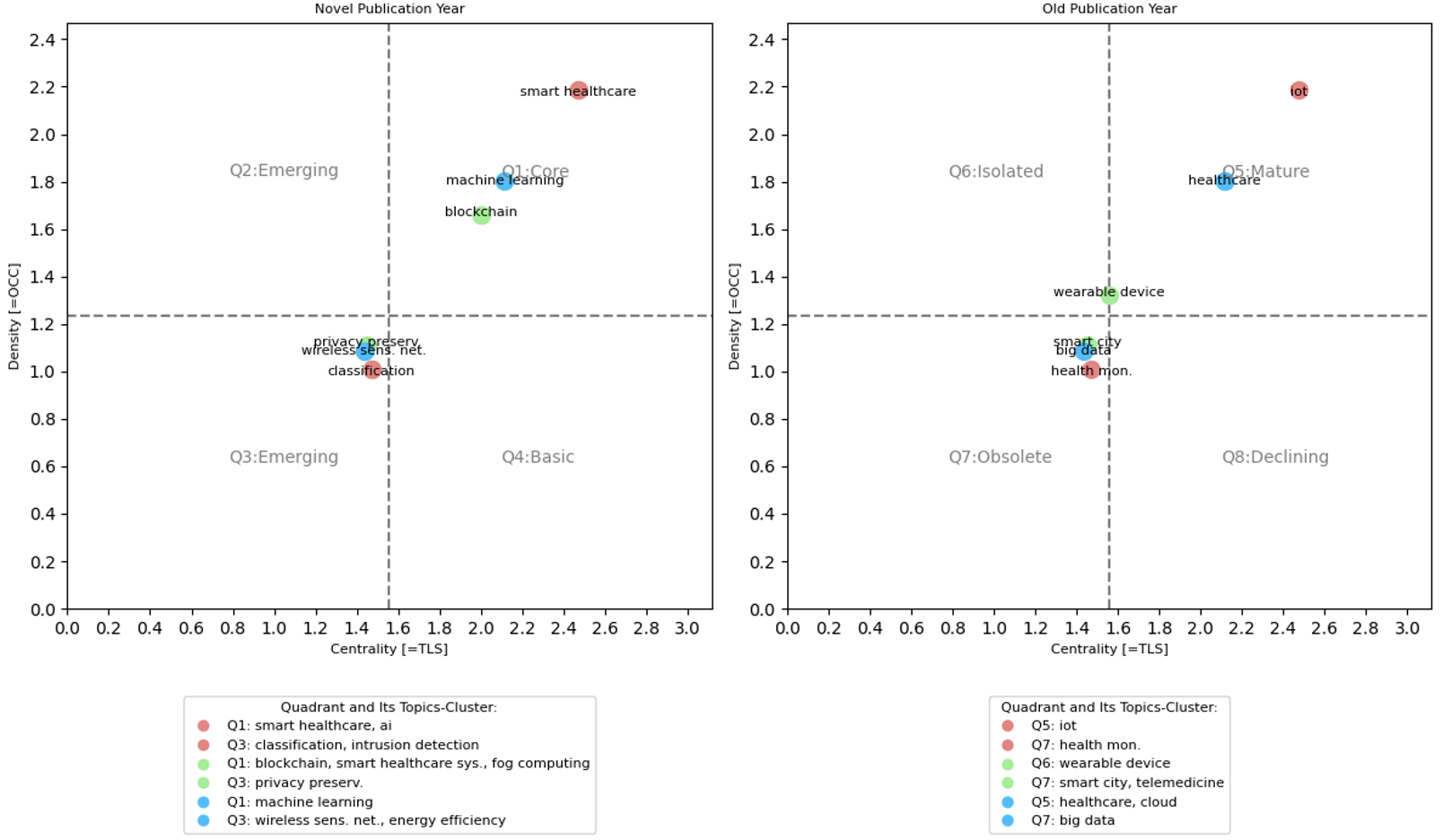} 
    \caption{Enhanced strategic diagram based on co-occurrence analysis of topics map}
    \label{fig:esd_diagram}
\end{figure}

In this study, the policy analysis was done by combining the main points of each cluster with technological trends, not only by using the literature based proof but also by referring to the Enhanced Strategic Diagram (ESD, \cref{fig:esd_diagram}) as a logical framework. The main emerging mature, and isolated quadrants were taken as examples because they refer to technologies having a potential policy impact that can be actually implemented, while going back to the overall patterns provides a cogent explanation of why these areas require attention (\cref{tab:policy_recommendations}). Such a method makes it possible that policy proposals can be followed from thematic understanding to practical and regulatory enforcement, especially for developing countries like Indonesia, where resource limitations, digital capability gaps, and infrastructure problems should be considered explicitly.

The main focus within Cluster 1, intelligent and secure AI-Driven IoT healthcare systems, is on the use of AI and smart healthcare to support both operational and clinical decision making. Technologies like smart healthcare and AI are central to this theme, allowing hospitals to establish AI governance frameworks, deploy validated clinical tools, and implement explainable AI models, ensuring clinical accountability. Regulators should complement these efforts by defining national AI standards, certification protocols, and auditing procedures to ensure safe and interoperable deployment \citep{chen2024, martens2024}. Emerging technologies such as intrusion detection and classification (Q3) further reinforce this theme, requiring hospitals to follow cybersecurity protocols, conduct network monitoring, and establish real-time threat detection mechanisms, while regulators enforce compliance standards and continuous auditing requirements \citep{amintoosi2022, bathalapalli2022}. Well-established IoT technologies (Q5) extend the theme into practical patient monitoring and asset tracking, where hospitals should integrate IoT into clinical operations in line with interoperability requirements, and regulators ensure adherence to national digital health policies and guidelines \citep{rahman2023, tuli2020}.

In Cluster 2, decentralized and privacy preserving digital health ecosystems, the overarching topic of discussion is around building trust, sharing data securely, and preserving privacy. Advanced technologies, such as blockchain, smart healthcare systems, and fog computing (Q1) are primary examples that show this topic vividly by allowing hospitals to secure health record management systems in a decentralized way, on the other hand, the regulators ought to come up with rules and frameworks for interoperability, patient consent, data ownership, and auditability \citep{fetjah2021, jamil2020}. Besides, privacy preserving technologies (Q3), especially federated learning and cryptography will need hospitals to have privacy preserving data governance, consent management, and collaborative analytics. Whereas, regulators should be the ones to enforce multi institutional collaboration guidelines and privacy compliance standards \citep{chen2024, chow2025}. Next, wearable devices (Q6) on their own are a means of remote patient monitoring where hospitals can connect wearable technologies through standardized clinical and data integration protocols meanwhile regulators can promote device safety, interoperability standards, and national digital health promotion policies \citep{gao2022, mohammed2023}.

Cluster 3, scalable cloud edge infrastructure for smart hospital ecosystems, focuses primarily on the main subject of building the most effective, durable, and data driven infrastructure in smart hospitals. Fundamental machine learning technologies (Q1) that assist us with this topic are the hospitals' operations namely through the implementation of predictive maintenance, optimization of clinical workflows, and analytics based on data. On the other hand, regulators have a role in setting performance indicators and standards for assessing the systems to ensure their effectiveness in guiding adoption \citep{iqbal2023, islam2020}. As for the new wireless sensor networks and energy saving technologies (Q3), they necessitate hospitals to practice resource optimization, energy management, and the implementation of sustainable IT. Besides, regulators should not only set energy efficiency requirements but also make available funded incentives for the development of sustainable healthcare infrastructures \citep{chanak2020, li2023, morovisconti2020}. Moreover, well established cloud computing and big data tools (Q5) support the capability to scale data management, ensure interoperability, and exchange information securely. Hospitals, in this case, should be the ones who bring together cloud edge setups with their clinical and operational systems. On the other hand, regulators need to create nationwide cloud health infrastructure programs as well as public private data sharing mechanisms that foster interoperability and scalability \citep{hoang2022, ji2024, nancy2022}.

\setlength{\LTpre}{0pt}
\begin{longtable}{|p{0.2\textwidth}|p{0.3\textwidth}|p{0.45\textwidth}|}
    \caption{Policy recommendation for each cluster/pattern with corresponding ESD quadrant}
    \label{tab:policy_recommendations} \\
    
    \hline
    \textbf{Cluster - Theme / Pattern} & \textbf{Key Technologies (ESD Quadrant)} & \textbf{Policy Recommendations} \\
    \hline
    \endfirsthead
    
    \multicolumn{3}{c}{{\bfseries \tablename\ \thetable{} -- continued from previous page}} \\
    \hline
    \textbf{Cluster - Theme / Pattern} & \textbf{Key Technologies (ESD Quadrant)} & \textbf{Policy Recommendations} \\
    \hline
    \endhead
    
    \hline \multicolumn{3}{|r|}{{Continued on next page...}} \\ \hline
    \endfoot
    
    \hline
    \endlastfoot
    
    1 - Intelligent and Secure AI Driven IoT Healthcare Systems & 
    \begin{itemize}[nosep, leftmargin=*, before=\vspace{-0.6\baselineskip}, after=\vspace{0pt}]
        \item Q1 (Core): Smart healthcare, AI
        \item Q3 (Emerging): Intrusion detection, classification
        \item Q5 (Mature): IoT
    \end{itemize} & 
    \textbf{Hospital:} Implement AI governance frameworks; deploy validated and explainable clinical AI systems; integrate real-time monitoring and cybersecurity mechanisms into clinical operations. \newline
    \textbf{Regulator:} Establish national AI standards; develop certification and audit frameworks; enforce IoMT cybersecurity compliance; ensure interoperability across digital health systems. \\
    \hline
    
    2 - Decentralized and Privacy Preserving Digital Health Ecosystems & 
    \begin{itemize}[nosep, leftmargin=*, before=\vspace{-0.6\baselineskip}, after=\vspace{0pt}]
        \item Q1 (Core): Blockchain, smart healthcare systems, fog computing
        \item Q3 (Emerging): Privacy-preserving technologies
        \item Q6 (Isolated): Wearable devices
    \end{itemize} & 
    \textbf{Hospital:} Adopt decentralized health record management; implement privacy-preserving data governance and federated learning; integrate wearable-enabled remote patient monitoring systems. \newline
    \textbf{Regulator:} Establish legal frameworks for data interoperability and patient consent; enforce multi-institutional privacy compliance and data governance standards; develop regulations for device safety, interoperability, and digital health adoption. \\
    \hline
    
    3 - Scalable Cloud Edge Infrastructure for Smart Hospital Ecosystems & 
    \begin{itemize}[nosep, leftmargin=*, before=\vspace{-0.6\baselineskip}, after=\vspace{0pt}]
        \item Q1 (Core): Machine learning
        \item Q3 (Emerging): Wireless sensor networks, energy efficiency
        \item Q5 (Mature): Cloud computing, big data
    \end{itemize} & 
    \textbf{Hospital:} Implement machine learning--driven predictive maintenance and clinical workflow optimization; apply data-driven clinical analytics; adopt energy-efficient resource management and sustainable IT deployment; enable scalable cloud--edge infrastructure integration. \newline
    \textbf{Regulator:} Establish national cloud health infrastructure initiatives; define interoperability and data exchange frameworks; enforce energy efficiency standards; provide funding incentives for sustainable and scalable healthcare infrastructure. \\
\end{longtable}

By explicitly connecting the key ideas of clusters with technological quadrants, policymaking, and the levels of implementation, this combined story creates a logical "red thread" from analysis to practical policies. It identifies hospitals, mainly in developing country settings, as the main actors who are able to transform themselves into smart hospitals by following well planned and stepwise strategies that are mindful of the use of resources, whereas regulators, as enablers and coordinators, provide the legal, technical, and institutional frameworks which guarantee interoperability, security, and standardization. Besides, the suggestions here are not only theoretically sound but also practically doable and allow the gradual uptake of innovations based on the level of infrastructure and the readiness of the organization without the need for a total overhaul. Thus, the framework which combines thematic patterns, the use of ESD for technological prioritization, and policy guidance targeted at different roles provides context sensitive and evidence based recommendations that help to narrow the gap between research knowledge and healthcare practice.

\section{Conclusion}
This study provides a structured and evidence-based synthesis of the evolving landscape of Intelligent and Secure Smart Hospital Ecosystems by integrating bibliometric analysis with the Scoping review and PAGER frameworks. The results identify three closely linked pillars that characterize the field, which are AI powered intelligent healthcare systems, decentralized and privacy preserving data ecosystems, and highly scalable cloud edge infrastructures. These pillars as a whole indicate a move away from isolated technology adoption towards a more integrated and ecosystem oriented paradigm, where intelligence, trust, and infrastructure are strongly interrelated. However, the study also points out that substantial progress has been made, especially in AI, blockchain, and cloud technologies, yet important limitations still exist with respect to interoperability, real world application, governance, and cross layer integration. Recognizing these underdeveloped areas offers a strong impetus for subsequent research, firmly pointing to the necessity of going beyond single innovations and adopting holistic and system level approaches.

From a practical and policy perspective, this research shows that advancing technology by itself is not enough if the governance and implementation strategies are not aligned with it. By combining the two bibliometric patterns and the Enhanced Strategic Diagram valuable insights can be drawn and policies can be formulated that are not only relevant to the region but also feasible especially to developing countries like Indonesia. It is advisable for hospitals to formalize AI governance, enhance cybersecurity and data management standards and employ scalable digital infrastructures. At the same time governments through legislation and establishment of interoperability standards and national digital health strategies, significantly contribute to this leading role. The joint emphasis on innovation, governance, and scalability suggests that a smart hospital ecosystem can deliver significantly improved healthcare services. This improvement, achieved through digital transformation, is sustainable and equitable only when technological and institutional dimensions are properly coordinated.

\section*{Author Contribution}
\noindent \textbf{AW} contributed to conceptualization, methodology, data curation, formal analysis, visualization, and writing of the original draft. \textbf{WMB} contributed to data curation and formal analysis support, and critical review and editing of the manuscript. \textbf{BH} contributed to data interpretation, validation, and review and editing. \textbf{CS} contributed to review and editing. All authors have read and agreed to the published version of the manuscript.


\end{document}